\documentclass[%
onecolumn,
 amsmath,amssymb,
 aps,
]{revtex4-2}

\usepackage{booktabs}
\usepackage{graphicx}
\usepackage{dcolumn}
\usepackage{bm}
\usepackage{hyperref}

\begin{document}

\title{Reconstructing Quantum Dot Charge Stability Diagrams with Diffusion Models}

\author{Vinicius Hernandes}
\email{v.hernandes@tudelft.nl}
\affiliation{QuTech and Kavli Institute of Nanoscience, Delft University of Technology, Delft, the Netherlands}

\author{Joseph Rogers}
\affiliation{QuTech and Kavli Institute of Nanoscience, Delft University of Technology, Delft, the Netherlands}

\author{Rouven Koch}
\affiliation{QuTech and Kavli Institute of Nanoscience, Delft University of Technology, Delft, the Netherlands}

\author{Thomas Spriggs}
\affiliation{QuTech and Kavli Institute of Nanoscience, Delft University of Technology, Delft, the Netherlands}

\author{Brennan Undseth}
\affiliation{QuTech and Kavli Institute of Nanoscience, Delft University of Technology, Delft, the Netherlands}

\author{Anasua Chatterjee}
\affiliation{QuTech and Kavli Institute of Nanoscience, Delft University of Technology, Delft, the Netherlands}

\author{Lieven M. K. Vandersypen}
\affiliation{QuTech and Kavli Institute of Nanoscience, Delft University of Technology, Delft, the Netherlands}

\author{Eliska Greplova}
\affiliation{QuTech and Kavli Institute of Nanoscience, Delft University of Technology, Delft, the Netherlands}

\date{\today}

\begin{abstract}
Efficiently characterizing quantum dot (QD) devices is a critical bottleneck when scaling quantum processors based on confined spins. Measuring high-resolution charge stability diagrams (or CSDs, data maps which crucially define the occupation of QDs) is time-consuming, particularly in emerging architectures where CSDs must be acquired with remote sensors that cannot probe the charge of the relevant dots directly. In this work, we present a generative approach to accelerate acquisition by reconstructing full CSDs from sparse measurements, using a conditional diffusion model. We evaluate our approach using two experimentally motivated masking strategies: uniform grid-based sampling, and line-cut sweeps. Our lightweight architecture, trained on approximately 9,000 examples, successfully reconstructs CSDs, maintaining key physically important features such as charge transition lines, from as little as 4\% of the total measured data. We compare the approach to interpolation methods, which fail when the task involves reconstructing large unmeasured regions. Our results demonstrate that generative models can significantly reduce the characterization overhead for quantum devices, and provides a robust path towards an experimental implementation.
\end{abstract}

\maketitle

\section{\label{sec:intro}Introduction}

\subsection{Quantum Dots and Charge Stability Diagrams}

\begin{figure}[t]
    \centering
    \includegraphics[width=\linewidth]{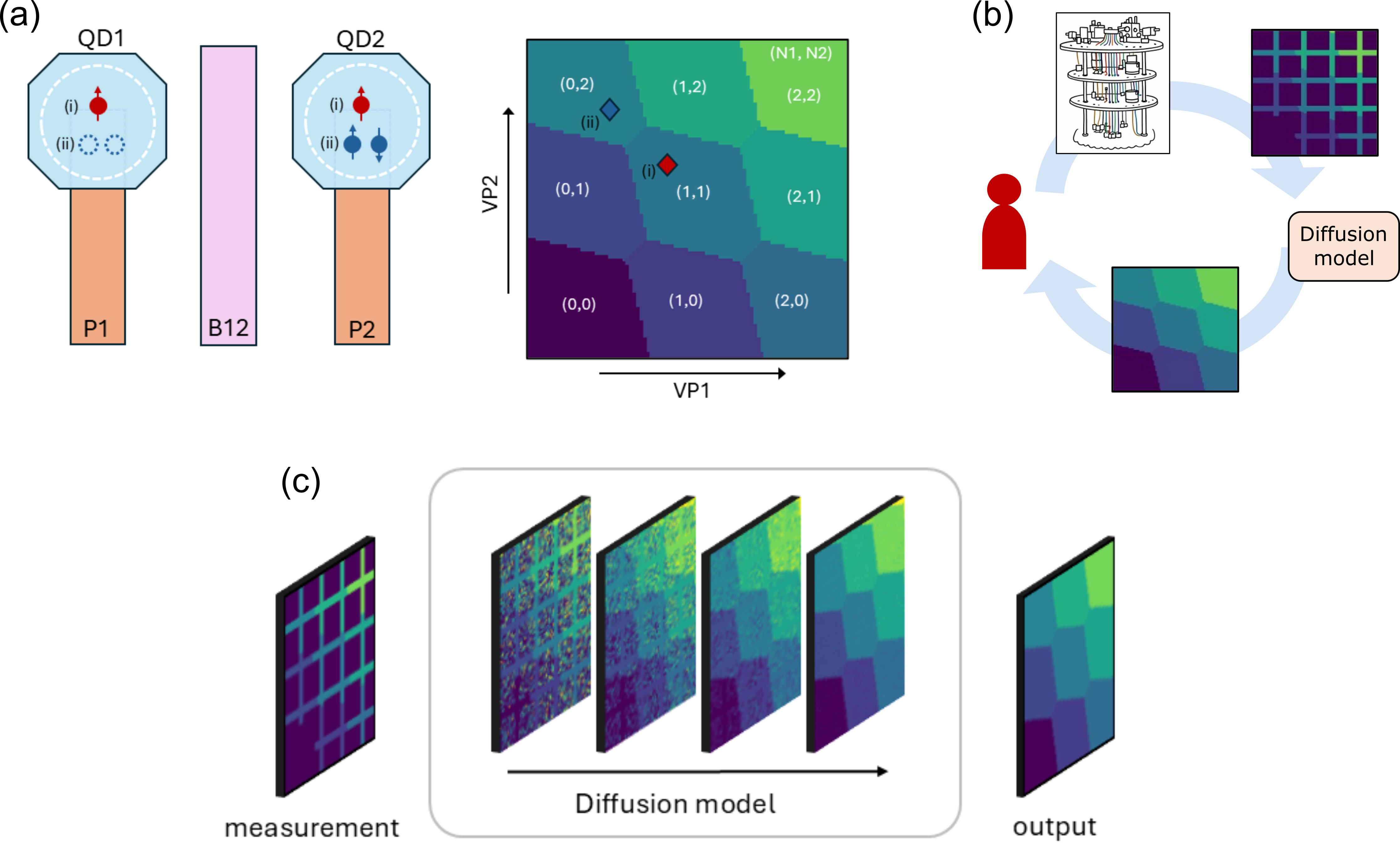}
    \caption{
    \textbf{Charge stability diagram of a (gate-defined) double quantum dot.}
    \textbf{(a)} Left: schematic of a double quantum dot.
    Right: (simulated) charge stability diagram as a function of two gate voltages. Colored regions correspond to fixed charge occupations $(N_1, N_2)$, while sharp transition lines indicate changes in the charge configuration. These transition lines encode the essential information required for device tuning (but are costly to acquire experimentally).
    (b) Measurement workflow.
    (c) Diffusion model reconstruction process.
    }
    \label{fig:main1}
\end{figure}

Semiconductor quantum dots hosting spin-based qubits are a promising platform for quantum computing due to their scalability and compatibility with established fabrication technologies~\cite{lossdivincenzo.57.120, spinqubits, petta2005coherent, veldhorst2015two, Vandersypen2017, Chatterjee2021, philips2022universal, Borsoi2024SharedControl}. In these systems, quantum information is encoded in the spin degree of freedom of confined charge carriers, and device operation relies on precise electrostatic control over multiple gate voltages. Achieving correct occupation and initialization, and maintaining suitable operating regimes, therefore requires accurate characterization of the device response across voltage space. As quantum processors scale toward larger quantum-dot arrays, efficient and automated characterization becomes a central experimental challenge.

Quantum-dot devices are typically characterized and tuned using the \emph{charge stability diagram} (CSD)~\cite{hanson2007spins}. Experimentally, CSDs are obtained by sweeping selected gate voltages and recording a charge-sensitive signal, such as the current through a nearby charge sensor, as illustrated in Figure~\ref{fig:main1}(a). The resulting two-dimensional map reveals regions of fixed charge occupation, separated by sharp transition lines corresponding to changes in the device configuration. These diagrams provide a compact representation of the underlying electrostatic landscape and form the basis of most tuning and calibration routines.

Device tuning typically follows an iterative, human-in-the-loop procedure in which gate voltages are adjusted, a CSD is measured and interpreted, and the process is repeated until a target operating regime is reached~\cite{ares2016sensitive}. This workflow also underpins automated tuning approaches, which extract physically meaningful parameters directly from features in the CSD~\cite{vanDiepen2018}. The most informative of these features are the charge transition lines, corresponding to charge loading, unloading, or inter-dot transfer, which delineate regimes suitable for qubit operation. Accurately identifying and preserving these transition lines is therefore central to efficient device control.

Despite their importance, charge stability diagrams can be expensive to acquire. While radio-frequency charge sensing \cite{vigneau2023probing} has significantly reduced acquisition times compared to traditional DC techniques, measurement time remains a limiting factor, particularly for tasks requiring extensive data collection such as automated tuning, statistical characterization, or machine learning applications \cite{carlsson2025automated}. In extended quantum dot systems where direct charge sensing is not feasible, the expectation value of a shuttled spin state may be used to infer charge stability regions~\cite{undseth2026weight}. However, the time required to acquire CSDs in this manner can easily exceed several minutes, as each pixel constitutes hundreds of measurement cycles. Combined with limited experimental access and device availability, this creates a practical bottleneck in device characterization and motivates the development of methods that reduce the experimental cost of generating large datasets for automated tuning and analysis, a need recently emphasized by the community \cite{zwolak2024data}.

Although CSDs are typically presented as two-dimensional images, they represent structured projections of a higher-dimensional control space involving many additional gate voltages and device parameters. In practice, experimentally measured CSDs are often noisy, sparsely sampled, or partially observed due to time and experimental constraints. These characteristics make CSDs a compelling example of structured scientific data, in which well-defined physical constraints coexist with incomplete or noisy measurements.

From a machine learning perspective, this setting naturally motivates data-efficient approaches that leverage known physical structure to reconstruct missing information and reduce measurement overhead. Such methods offer a pathway toward accelerating device characterization while preserving the physically relevant features required for quantum-dot tuning.

\subsection{Machine Learning Based Automation}

Machine learning  has been extensively applied to automate the tuning routine of quantum dots, with approaches ranging from discriminative methods used to classify charge transitions to genetic algorithms for optimizing experimental parameters such as pulse amplitudes and ramp times ~\cite{kalantre2019machine,lennon2019efficiently,durrer2020automated,zwolak2020autotuning,darulova2020autonomous,czischek2021miniaturizing,nguyen2021deep,bucko2023automated,schuff2024fully, katiraee2025unified}. While these methods have proven valuable for automating tuning pipelines, they typically require long measurement times to acquire full CSDs.

    We take a different approach by treating CSD acquisition as a partial observation problem. Rather than measuring complete voltage ranges, we propose using a generative model to reconstruct missing measurements from strategically-designed sparse scans. This workflow enables a speed-up compared to the standard  full measurement case: experimentalists can measure fewer voltage points and use a diffusion model to reconstruct the complete diagram, freeing up the device more quickly for subsequent measurements or different experimental tasks.

Our approach is motivated by the observation that CSDs present strong structural features that can be learned and exploited for reconstruction, including approximately linear transition lines,  charge regions defined by  characteristic geometries, and familiar noise patterns. Unlike classical interpolation methods, that assume local smoothness, diffusion models can capture these structural priors when trained on a dataset of complete CSDs.

The final goal of our reconstruction is not necessarily pixel fidelity, but accurate recovery of the most experimentally relevant information in a CSD: charge transition lines. These are chosen as they encode the physics required to tune the device. Therefore, our evaluation prioritizes correct identification and positioning of transition lines, ensuring that real lines are detected and not spuriously introduced, over general image quality metrics

\begin{figure}[t]
  \centering
  \includegraphics[width=\linewidth]{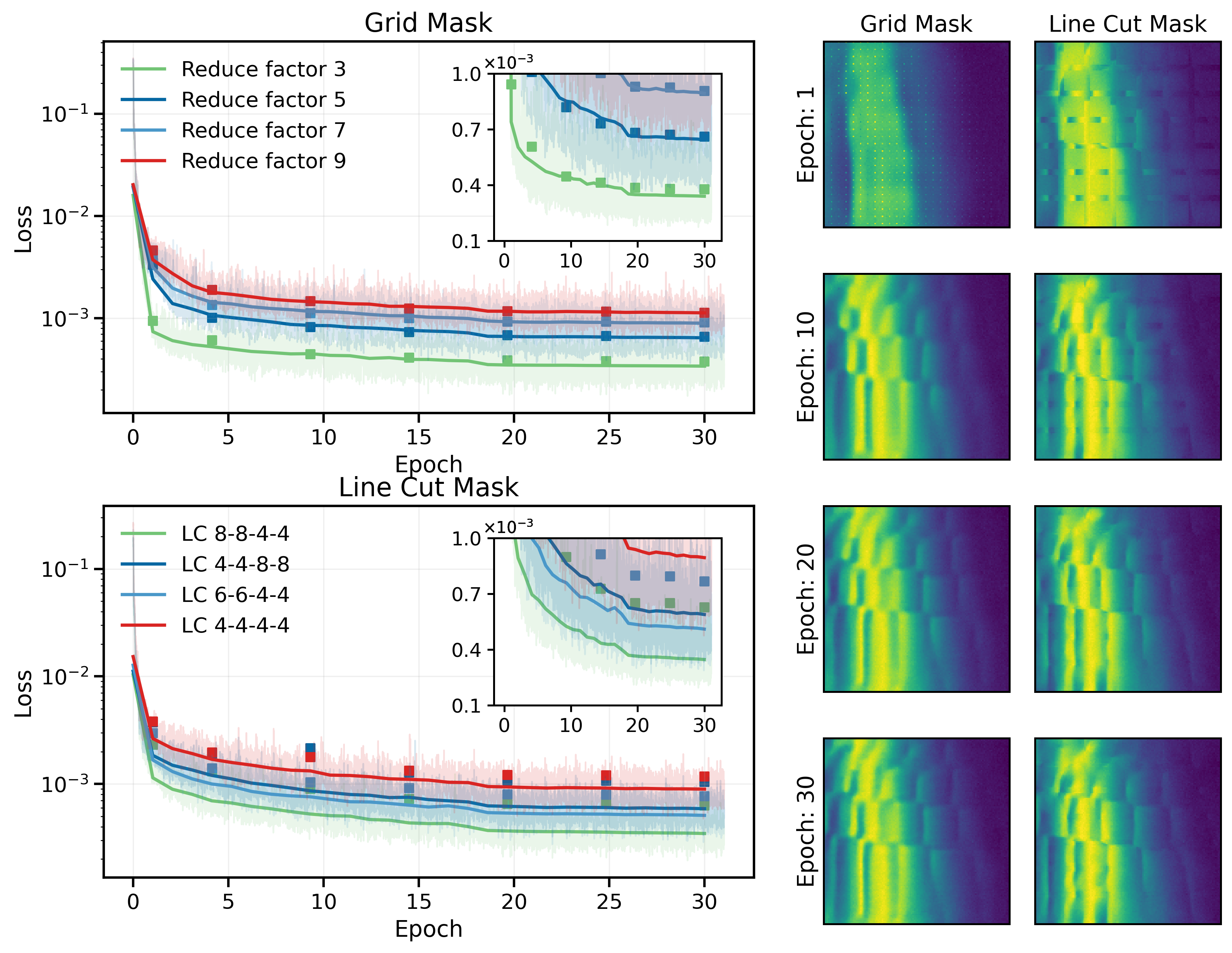}
  \caption{Training loss over 30 epochs for grid masks (top) and line-cut masks (bottom) at different sparsity levels using 140 diffusion steps. The solid line shows the average training loss per epoch, while the shaded region indicates the training loss variation across training steps (mini-batches), and the square markers show validation loss for selected epochs. Reconstructions at the right side correspond to epochs 1, 10, 20, and 30 for both masking strategies.
}
  \label{loss}
\end{figure}

\subsection{Contributions}

This work makes the following contributions:

\begin{enumerate}
    \item \textbf{Domain-adapted diffusion architecture}: We use a conditional diffusion model with a lightweight U-Net backbone, designed for fast inference on limited scientific data. Unlike traditional large-scale diffusion models, which are trained on millions of examples, our architecture is fit to the scarce-data regime, while maintaining the ability to learn global structural patterns. The model allows for millisecond inference times, making it suitable for experimental feedback loops.
    \item \textbf{Physically meaningful evaluation metrics}: We employ an evaluation routine that reflects the accuracy in reconstructing physically relevant information. By extracting ridges and edges from the original and reconstructed CSDs, which highlight charge transition lines, we directly assess how well tuning-relevant structure is preserved.  These results are  evaluated alongside standard pixel-based image quality metrics.
    \item \textbf{Measurement strategy and experimental feasibility}: We systematically compare two experimentally relevant measurement routines, grid masks (uniform subsampling), and line-cut masks along vertical and horizontal axes. The axis-aligned cuts provide a standardized pattern to evaluate how leaving large regions unmeasured affects reconstruction, though experimental sweeps need not be restricted to these directions. We study these strategies across sparsity levels and diffusion steps, revealing trade-offs between measurement density, reconstruction accuracy, and inference time, and providing practical guidance for choosing measurement protocols in different experimental scenarios. Within this trade-off, we show that measurement schemes can be designed to significantly reduce acquisition time while maintaining accurate CSD reconstructions with faithfully preserved charge transition lines.
\end{enumerate}

\section{Methods}

\subsection{Dataset}

We use the Delft Charge Stability Diagram Dataset \cite{de_snoo_2025_16363472}, a curated collection of CSDs measured over a three-year period, across multiple quantum-dot devices and experimental groups at QuTech/TU Delft. The dataset is a subset of a larger body of approximately 120,000 CSD measurements, with diverse structures, features, and noise-levels.

To create a high-quality training set, a filtering approach to the original CSD measurement is applied. A small subset (4000) of CSDs are manually labeled by experts as either ``clean'' (well-defined transition lines, low noise) or ``noisy'' (unclear features, high noise). A classification model was then trained on these labeled examples and used to score the remaining CSDs. By setting an appropriate threshold, a filtered dataset is obtained, containing 9850 CSDs that exhibit clear charge transition features suitable for reconstruction tasks. For more details regarding the dataset used, refer to \cite{de_snoo_2025_16363472}.

From this curated dataset, we manually selected 20 diverse CSDs for our test set, choosing examples with varying transition line geometries and noise characteristics. This test set is used solely for final evaluation of the trained models, and the CSD examples presented in this work are all taken from it. Of the remaining samples, we allocate 90\% for training and 10\% for validation and hyperparameter tuning.

The dataset size, while small compared to those used for standard diffusion models training, is characteristic of the kinds of scientific domains where data acquisition is expensive and time-consuming \cite{xu2023small, dietrich2025scientific, khader2023denoising}. Our model has to learn from limited examples while generalizing to diverse CSD structures.

\subsection{Measurement Routine: Masks and Measurement Density}

In order to simulate realistic experimental measurement protocols, we design two masking patterns that correspondent to two distinct voltage scanning strategies.

\textbf{Grid Mask}: Measures 1 out of every $n$ voltage points in scanning each direction, when compared to the original measurement. This directly corresponds to multiplying the voltage step size between every measurement by $n$, which we call the \textit{reduce factor}. For example, a reduce factor of 3 measures every third point vertically and horizontally, which corresponds to measuring $1/9^\text{th}$ of the original data. This strategy is straightforward to implement experimentally in current settings (simply by multiplying the voltage step by $n$) and has the advantage that masked (unmeasured) points are never far from measured points, providing  local context throughout the image. This makes the reconstruction task easier since the diffusion model always has nearby conditioning information. We test the grid mask approach with reduce factors 3, 5, 7, and 9, corresponding to measuring approximately 11\%, 4\%, 2\%, and 1\% of the original data, respectively. These parameters progressively reduce the available measurement context for the diffusion model, increasing reconstruction difficulty while reducing measurement time.

\textbf{Line Cut Mask}: In this approach,  horizontal and vertical line scans across the voltage space are measured. We define the \textit{line-cut factor} as $(N_h, N_v, T_h,T_v)$, where $N_h$ and $N_v$ are the number of horizontal and vertical sweeps, and $T_h$ and $T_v$  the pixel thickness (or voltage step in the experiment) of each sweep. For example, (8, 8, 4, 4) performs 8 horizontal and 8 vertical line-cuts, each 4 pixels wide. Line-cut masks are more difficult for reconstruction tasks, since they leave large regions of missing data, far from the conditional context. While this strategy may be useful for initial exploratory sweeps, the measurement savings compared to a grid mask are limited. In practice, experimentalists could perform sweeps along arbitrary directions (rays \cite{zwolak2021ray}) in the voltage space; the axis-aligned line cuts considered here provide a standardized benchmark to test model performance on challenging, nonuniform sampling patterns.

We test the line-cut approach with line-cut factors (8,8,4,4), (6,6,4,4), (4,4,8,8), and (4,4,4,4), which, for the image size considered in this work, 128×128, correspond to measuring approximately 44\%, 34\%, 44\%, and 23\% of the original data respectively. Note that (8,8,4,4) and (4,4,8,8) have identical density but different spatial distributions: the former has more but thinner line-cuts, while the latter has fewer but thicker scans.

\section{Diffusion Model}

\subsection{Conditional Reconstruction}

We employ a conditional denoising diffusion probabilistic model (DDPM) \cite{solh-dickstein2015deep, ho2020denoising}, adapted for sparse measurement reconstruction \cite{lugmayr2022repaint}. The core principle of  diffusion models is to learn a gradual denoising procedure: starting from pure Gaussian noise, the model iteratively removes noise over  $T$ time steps to generate a clean image, guided by the observed measurements \cite{dhariwal2021diffusion}.

Our approach builds on a recent work on diffusion-based image inpainting with internal learning, in which the model learns with datasets as small as one example \cite{cherel2024diffusion}. In this work, however, we train the model with approximately 9,000 examples, allowing  our model to learn general structural patterns across diverse CSDs while remaining computationally efficient.

We employ explicit spatial conditioning by concatenating the noisy input with the masked measurement, the binary mask, and the time embedding.

\subsection{Architecture}

Following standard practice in diffusion models, we map the discrete time step $t$ to a 16-dimensional vector using a sinusoidal encoding, 
\begin{equation}
    emb(t) = [\sin(\omega_1t),\cos(\omega_1t),\ldots,\sin(\omega_8t),\cos(\omega_8t)],
\end{equation}
which is then refined  through a two-layer MLP (16 $\rightarrow$ 16 $\rightarrow$ 16 dimensions with GELU activation), and finally spatially broadcast to create a 16-channel time map that the U-Net uses to modulate the denoising intensity. This embedding allows the model to distinguish between early steps, with high noise, to late steps, with less noise and more fine details. 

The core of our architecture is a multi-scale U-Net \cite{ronneberger2015u}, which operates at three distinct stages:
\begin{enumerate}
    \item \textbf{Input space projection}: A projection block receives a 19-channel tensor consisting of the noisy input $x_t$, the sparse measurement $y$, the binary mask $M$, and the 16-channel broadcasted time embedding. This block uses a $3\times3$ convolutional layer to map the input into a 32-channel latent representation.
    \item \textbf{Hierarchical encoder-decoder}: The model processes information across 3 levels. At each level, an encoder block applies two $3\times3$ convolutions with ReLU activation followed by a $2\times2$ max-pooling. This cuts the input resolution in half. Afterwards, upsampling is performed using bilinear interpolation, bringing the image to previous level size, which is then concatenated with the features saved from the encoder via skip-connections. 
    \item  \textbf{Image reconstruction}: A final $3\times3$ convolution is applied, mapping back the 32 latent channels to a single channel image.

\end{enumerate}

The multi-scale design of the U-Net is particularly fit to the physical structure of charge stability diagrams. By processing the data across three hierarchical levels, the model captures global context, allowing it to infer the CSD geometry and fill large gaps between measured points. Simultaneously, skip-connections help preventing the loss of transition lines during downsampling.

Our complete model contains approximately 160 thousand parameters, significantly smaller than usual diffusion models used for natural images. This small size allows for fast inference times (sub-second on GPU), and efficient training on limited data. This specific architecture has shown great results in the domain of scarce, scientific data image reconstruction. 

\subsection{Training}

At each training iteration, we sample a clean CSD from the training set, add noise according to the diffusion schedule, and train the model to predict the denoised image, given the noisy image, sparse measurements, mask, and timestep. The loss function is simply the mean squared error between the original and the predicted image. 

After systematic validation experiments, we settled on the following hyperparameters: diffusion steps varying between 20, 60, 100, and 140, to assess the trade-off between inference time and accuracy, a linear noise schedule increasing from $\beta=0.0001$ to $\beta=0.02$, the Adam optimizer with learning rate $3\times10^{-4}$, and 30 training epochs.

Figure \ref{loss} shows the training loss over 30 epochs, and validation loss for selected checkpoints, for the grid mask (top panel) and line-cut mask (middle panel), and different sparsity levels, for the model using 140 diffusion steps. The loss converges around 20 epochs, indicating that longer training may not significantly improve performance. We save model checkpoints every 5 epochs to analyze how reconstruction quality improves with training. We show examples of reconstructed images for both masking strategies next to the loss curves, for epochs 1, 10, 20, and 30. The loss evolution is reflected in these results: at epoch 1 the image remains highly noisy; by epoch 10, mask-induced artifacts are still visible; by epoch 20 the reconstruction is largely clean, with only minor detail refinements observed up to epoch 30. It is also visible that there is a higher discrepancy between training and validation loss for the case of line-cut masks, compared to that of grid masks, showing that models trained on grid masks achieve better generalization.

 \section{Baseline Methods}

We compare our diffusion-based approach to three standard interpolation techniques used to reconstruct sparse measurements:

\begin{enumerate}
    \item \textbf{Linear Grid Interpolation}. Computes piecewise linear interpolation from the measured points. Interpolated values remain bounded by the values at neighboring data points, producing locally conservative reconstructions.
    \item \textbf{Inverse Distance Weighting (IDW)}. Computes interpolated values as a weighted average of nearby measured points, where weights are inversely proportional to distance raised to a power. We use k=8 nearest neighbors and power p=2. 
    \item \textbf{Biharmonic Interpolation}. Fits a thin-plate spline that minimizes bending energy, producing maximally smooth interpolations.
\end{enumerate}

These methods assume local smoothness, and do not incorporate prior knowledge about global structure of CSDs or typical noise patterns, a limitation that suggests promising directions for experimental follow-up, such as exploiting learned global structure to guide adaptive measurement. By focusing on diffusion versus classical interpolation, we isolate the key question: whether a data-driven generative model can provide advantages over classical interpolation approaches. While other learning-based reconstruction models (i.e. autoencoder, CNN, conditional GAN) could also be considered, including them would go beyond the intended scope of this study.

\section{Evaluation Metrics}

We evaluate reconstruction quality using two complementary sets of metrics that capture different aspects of the performance.

\textbf{Image metrics}. These standard metrics work on pixel-level accuracy and assess overall image quality without considering physical structure. We compute the \textit{root normalized mean squared error} (RNMSE), \textit{peak signal-to-noise ratio} (PSNR), and \textit{structural similarity index} (SSIM). While these metrics provide a standard basis for comparison, they are not sufficient for our domain: for the case considered here transition lines occupy a small portion of the total CSD, for which a reconstruction could achieve low error by accurately interpolating noise and background regions while completely missing the transition lines, the main experimentally relevant features.

\textbf{Structure aware metrics}. To assess whether reconstructions preserve physically meaningful structure, we extract charge transition lines from both ground truth and predicted CSDs using two complementary methods: edge extraction using the Canny algorithm \cite{canny2009computational}, and ridge detection using the Frangi filter \cite{frangi1998multiscale} and a threshold to produce a binary map. Ridges can be used to identify bright linear structures at multiple scales, and edges used to highlight regions of rapid intensity change. Reconstruction quality is then evaluated by computing three comparison metrics. \textit{Intersection-over-Union} (IoU) measures the overlap between true and predicted transition lines, with 1 indicating perfect overlap and 0 indicating no overlap. \textit{F1 Score} balances precision (avoiding false transitions) and recall (avoiding missed transitions). \textit{Hausdorff Distance} measures the maximum distance from any point in one feature set to the nearest point in the other, capturing worst-case localization error.

\begin{figure}[t]
  \centering
  \includegraphics[width=\linewidth]{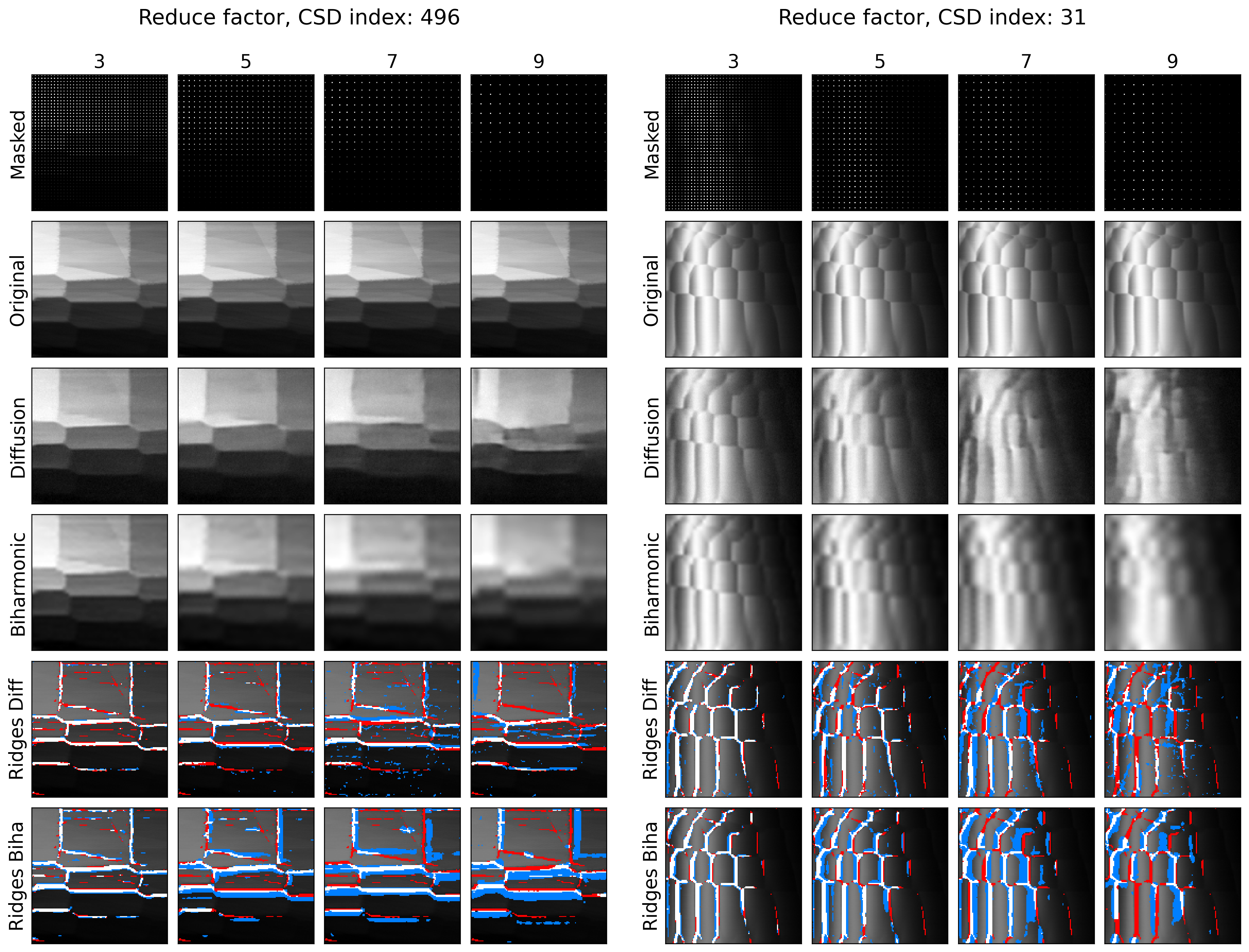}
  \caption{Reconstruction results for the grid mask strategy, presented for 140 diffusion steps and a selected example from the test set. Columns show different reduce factors. From top to bottom, the rows show: masked measurement, original full measurement, diffusion reconstruction, biharmonic interpolation reconstruction, ridges detection for diffusion model compared to original and ridges detection for biharmonic interpolation compared to original. Predicted ridges matching original are shown in white, ridges that are detected by the prediction but not the original are shown in blue, ridges that are present in the original example but missing in the prediction are shown in red. }
  \label{comparison_grid_mask}
\end{figure}

\section{Results}

\subsection{Grid Mask}

The performance of our reconstruction framework for the grid mask strategy highlights the advantage of a learned prior over classical interpolation. As shown in Table \ref{table1}, the diffusion model consistently matches or exceeds the performance of all baselines on structure aware metrics. While pixel-based metrics like PSNR are often similar to biharmonic interpolation, visual inspection shows that interpolation tends to blur the sharp transition lines with increasing the reduce factor (Figure \ref{comparison_grid_mask}).

When analyzing the quantitative results, it is noticeable that higher sparsity returns sub-optimal results, especially for reduce factor $\geq$ 7. However, compared to baselines, the diffusion model maintains faithfully reconstruction of ridges, as shown in Figure \ref{comparison_grid_mask}. Interestingly, pixel-level metrics (SSIM, PSNR) sometimes favor interpolation at these high sparsity levels. This occurs because interpolation generates smooth images that match noise and background characteristics well, while the diffusion model attempts to reconstruct sharp transitions that may not align perfectly with ground truth when conditioning information is very sparse. For our application, the superior structure metrics of diffusion are more important than pixel-wise smoothness.

\begin{figure}[t]
  \centering
  \includegraphics[width=\linewidth]{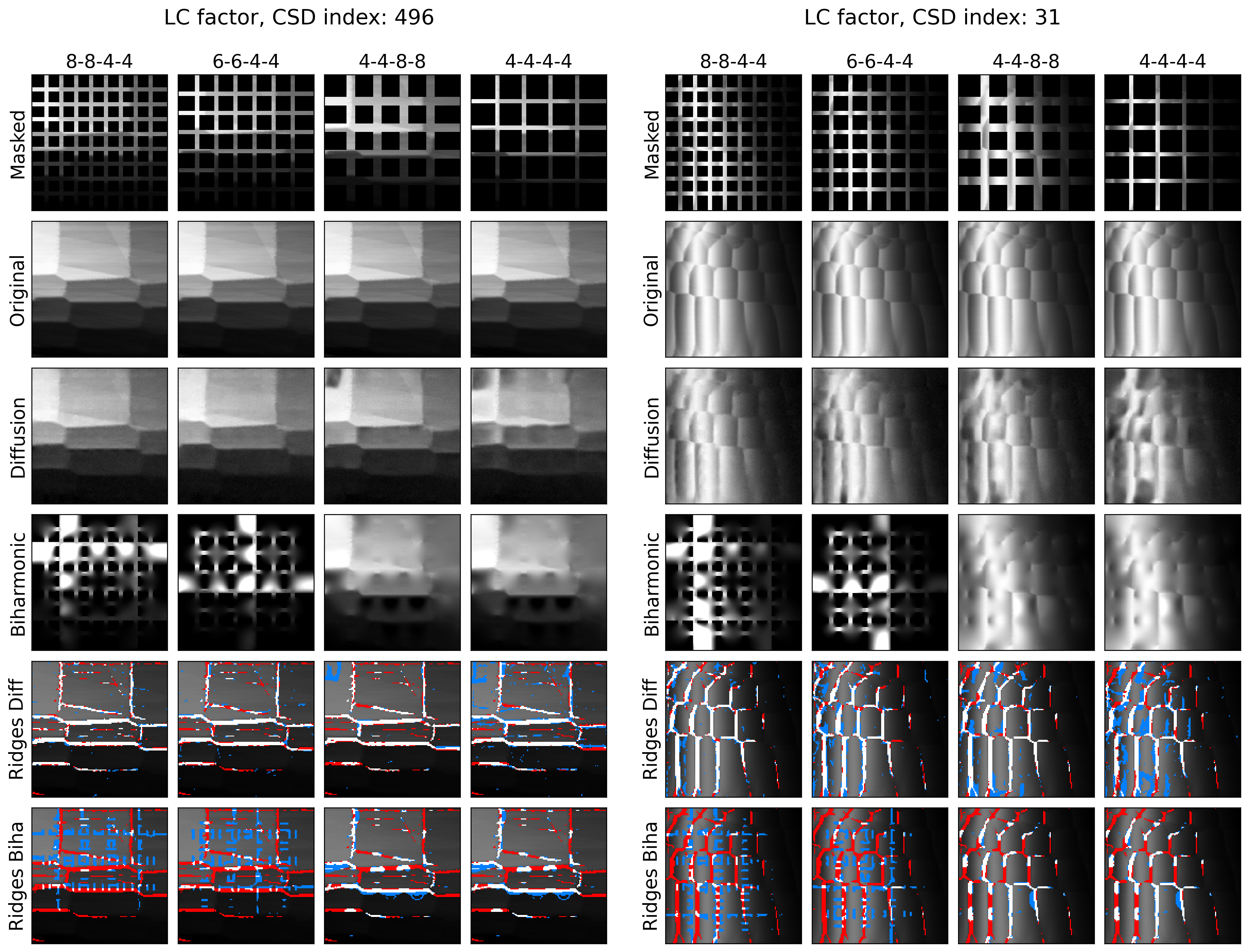}
  \caption{Reconstruction results for the line-cut mask strategy, presented for 140 diffusion steps and a selected example from the test set. Columns show different line-cut factors. From top to bottom, the rows show: masked measurement, original full measurement, diffusion reconstruction, biharmonic interpolation reconstruction, ridges detection for diffusion model compared to original (in white predicted ridges match original, in blue ridges are detected by the prediction but not the original, and in red ridges are present in the original example but missing in the prediction), and ridges detection for biharmonic interpolation compared to original.}
  \label{comparison_lc_mask}
\end{figure}

\subsection{Line Cut Mask}

Reconstructing CSDs from line-cut (LC) masks present a significantly more difficult task due to the large unobserved regions between sweeps. Metrics in Table \ref{table2} reveal that classical interpolation techniques effectively fail in this regime, with RNMSE above 0.6 and PSNR below 10 dB for most cases. These methods lack knowledge of global structure, which makes it difficult to accurately fill regions far from measured context. 

Diffusion models, on the other hand, show the ability to reconstruct the CSDs even for this more challenging task. Both pixel-wise and structural metrics in Table \ref{table2} show high accuracy. This is reflected in Figure \ref{comparison_lc_mask}, where visual inspection shows that reconstruction is successful, even when comparing transition lines via ridges overlay. However, the model reaches a limit for the extremely sparse configuration of line-cut factor (4,4,4,4), for which the metrics are significantly worse. 

Comparing configurations LC (8,8,4,4) and LC (4,4,8,8) provides crucial insight. Despite both measuring identical densities, the former achieves substantially better reconstruction (IoU = 0.574 vs 0.473, F1 = 0.720 vs 0.628). This reveals that spatial coverage, having many sweeps even if thin, is more valuable than measuring fewer locations more densely. The model benefits from seeing transition structure at more spatial locations rather than simply having more data along fewer lines.

\begin{table}[h]
    \label{table1}
  \caption{Reconstruction metrics at diffusion steps = 140. Mean $\pm$ std over test set.}
  \centering
  \begin{tabular}{lcccccc}
    \toprule
    Method
    & rNMSE $\downarrow$ & PSNR $\uparrow$ & SSIM $\uparrow$ & IoU $\uparrow$ & F1 $\uparrow$ & Haus. $\downarrow$ \\
    \midrule
    \multicolumn{7}{l}{\textbf{Reduce factor 3}} \\
    Diffusion
    & \textbf{0.054$\pm$0.031} & \textbf{31.89$\pm$4.18} & 0.846$\pm$0.074 & \textbf{0.529$\pm$0.172} & \textbf{0.676$\pm$0.155} & 21.74$\pm$8.35 \\
    Biharmonic
    & 0.059$\pm$0.034 & 31.15$\pm$4.45 & \textbf{0.875$\pm$0.088} & 0.492$\pm$0.124 & 0.651$\pm$0.112 & \textbf{20.07$\pm$7.79} \\
    Griddata (linear)
    & 0.061$\pm$0.033 & 30.78$\pm$4.28 & 0.874$\pm$0.085 & 0.478$\pm$0.116 & 0.639$\pm$0.109 & 22.39$\pm$8.45 \\
    IDW ($k{=}8$, $p{=}2$)
    & 0.065$\pm$0.031 & 30.11$\pm$4.06 & 0.868$\pm$0.081 & 0.457$\pm$0.106 & 0.621$\pm$0.099 & 23.91$\pm$8.44 \\
    \midrule
    \multicolumn{7}{l}{\textbf{Reduce factor 5}} \\
    Diffusion
    & \textbf{0.079$\pm$0.037} & \textbf{28.39$\pm$3.85} & 0.781$\pm$0.090 & \textbf{0.399$\pm$0.152} & \textbf{0.555$\pm$0.152} & \textbf{24.14$\pm$9.04} \\
    Biharmonic
    & 0.086$\pm$0.037 & 27.57$\pm$3.86 & 0.797$\pm$0.109 & 0.323$\pm$0.095 & 0.481$\pm$0.107 & 28.30$\pm$10.10 \\
    Griddata (linear)
    & 0.087$\pm$0.036 & 27.48$\pm$3.78 & \textbf{0.799$\pm$0.102} & 0.278$\pm$0.075 & 0.429$\pm$0.093 & 28.60$\pm$9.04 \\
    IDW ($k{=}8$, $p{=}2$)
    & 0.090$\pm$0.035 & 27.15$\pm$3.72 & 0.786$\pm$0.101 & 0.287$\pm$0.076 & 0.441$\pm$0.090 & 28.69$\pm$9.76 \\
    \midrule
    \multicolumn{7}{l}{\textbf{Reduce factor 7}} \\
    Diffusion
    & \textbf{0.105$\pm$0.044} & \textbf{25.78$\pm$3.68} & 0.722$\pm$0.099 & \textbf{0.315$\pm$0.134} & \textbf{0.465$\pm$0.145} & \textbf{24.30$\pm$8.92} \\
    Biharmonic
    & 0.110$\pm$0.040 & 25.43$\pm$3.92 & 0.733$\pm$0.121 & 0.244$\pm$0.072 & 0.388$\pm$0.090 & 27.36$\pm$8.68 \\
    Griddata (linear)
    & 0.108$\pm$0.038 & 25.54$\pm$3.82 & \textbf{0.747$\pm$0.115} & 0.194$\pm$0.068 & 0.320$\pm$0.095 & 30.28$\pm$9.24 \\
    IDW ($k{=}8$, $p{=}2$)
    & 0.110$\pm$0.037 & 25.36$\pm$3.68 & 0.727$\pm$0.110 & 0.222$\pm$0.059 & 0.359$\pm$0.078 & 27.52$\pm$9.06 \\
    \midrule
    \multicolumn{7}{l}{\textbf{Reduce factor 9}} \\
    Diffusion
    & 0.144$\pm$0.055 & 23.04$\pm$3.85 & 0.659$\pm$0.116 & \textbf{0.223$\pm$0.087} & \textbf{0.357$\pm$0.114} & \textbf{26.97$\pm$9.25} \\
    Biharmonic
    & 0.137$\pm$0.053 & 23.54$\pm$4.01 & 0.678$\pm$0.131 & 0.196$\pm$0.061 & 0.324$\pm$0.084 & 29.68$\pm$10.23 \\
    Griddata (linear)
    & \textbf{0.132$\pm$0.048} & \textbf{23.76$\pm$3.83} & \textbf{0.696$\pm$0.126} & 0.120$\pm$0.054 & 0.211$\pm$0.087 & 35.96$\pm$11.00 \\
    IDW ($k{=}8$, $p{=}2$)
    & 0.134$\pm$0.046 & 23.59$\pm$3.64 & 0.671$\pm$0.123 & 0.177$\pm$0.046 & 0.298$\pm$0.068 & 25.95$\pm$7.88 \\

    \bottomrule
  \end{tabular}
\end{table}

\begin{table}[h]
    \label{table2}
  \caption{Reconstruction metrics at diffusion steps = 140. Mean $\pm$ std over test set.}
  \centering
  \begin{tabular}{lcccccc}
    \toprule
    Method
    & rNMSE $\downarrow$ & PSNR $\uparrow$ & SSIM $\uparrow$ & IoU $\uparrow$ & F1 $\uparrow$ & Haus. $\downarrow$ \\
    \midrule
        \multicolumn{7}{l}{\textbf{LC 8-8-4-4}} \\
    Diffusion
    & \textbf{0.054$\pm$0.022} & \textbf{31.47$\pm$3.33} & \textbf{0.870$\pm$0.038} & \textbf{0.574$\pm$0.139} & \textbf{0.720$\pm$0.119} & \textbf{20.63$\pm$7.00} \\
    Biharmonic
    & 0.673$\pm$0.041 & 8.99$\pm$1.00 & 0.149$\pm$0.042 & 0.073$\pm$0.024 & 0.135$\pm$0.041 & 27.16$\pm$5.95 \\
    Griddata (linear)
    & 0.746$\pm$0.031 & 8.08$\pm$0.95 & 0.118$\pm$0.036 & 0.074$\pm$0.030 & 0.136$\pm$0.053 & 30.16$\pm$8.89 \\
    IDW ($k{=}8$, $p{=}2$)
    & 0.734$\pm$0.012 & 8.22$\pm$1.01 & 0.126$\pm$0.039 & 0.098$\pm$0.028 & 0.178$\pm$0.047 & 26.62$\pm$7.29 \\
    \midrule
    \multicolumn{7}{l}{\textbf{LC 6-6-4-4}} \\
    Diffusion
    & \textbf{0.079$\pm$0.035} & \textbf{28.45$\pm$4.21} & \textbf{0.839$\pm$0.045} & \textbf{0.508$\pm$0.162} & \textbf{0.658$\pm$0.148} & \textbf{23.07$\pm$7.2}1 \\
    Biharmonic
    & 0.757$\pm$0.047 & 7.96$\pm$1.01 & 0.101$\pm$0.027 & 0.067$\pm$0.022 & 0.125$\pm$0.039 & 29.39$\pm$6.45 \\
    Griddata (linear)
    & 0.807$\pm$0.029 & 7.40$\pm$1.11 & 0.080$\pm$0.024 & 0.071$\pm$0.020 & 0.132$\pm$0.035 & 30.87$\pm$7.51 \\
    IDW ($k{=}8$, $p{=}2$)
    & 0.816$\pm$0.018 & 7.29$\pm$1.05 & 0.076$\pm$0.025 & 0.091$\pm$0.023 & 0.165$\pm$0.039 & 28.88$\pm$7.35 \\
    \midrule
    \multicolumn{7}{l}{\textbf{LC 4-4-8-8}} \\
    Diffusion
    & 0.170$\pm$0.056 & 21.48$\pm$4.07 & \textbf{0.812$\pm$0.045} & \textbf{0.473$\pm$0.149} & \textbf{0.628$\pm$0.153} & \textbf{23.87$\pm$5.61} \\
    Biharmonic
    & \textbf{0.162$\pm$0.065} & \textbf{22.22$\pm$4.04} & 0.755$\pm$0.097 & 0.314$\pm$0.053 & 0.476$\pm$0.060 & 28.13$\pm$6.75 \\
    Griddata (linear)
    & 0.171$\pm$0.060 & 21.51$\pm$3.41 & 0.709$\pm$0.099 & 0.205$\pm$0.059 & 0.337$\pm$0.083 & 28.75$\pm$8.38 \\
    IDW ($k{=}8$, $p{=}2$)
    & 0.180$\pm$0.057 & 20.94$\pm$2.92 & 0.728$\pm$0.089 & 0.252$\pm$0.076 & 0.397$\pm$0.099 & 27.11$\pm$8.06 \\
    \midrule
    \multicolumn{7}{l}{\textbf{LC 4-4-4-4}} \\
    Diffusion
    & 0.266$\pm$0.086 & 17.58$\pm$4.04 & 0.693$\pm$0.079 & \textbf{0.355$\pm$0.159} & \textbf{0.505$\pm$0.176} & 28.34$\pm$11.10 \\
    Biharmonic
    & \textbf{0.162$\pm$0.065} & \textbf{22.22$\pm$4.04} & \textbf{0.755$\pm$0.097} & 0.314$\pm$0.053 & 0.476$\pm$0.060 & 28.13$\pm$6.75 \\
    Griddata (linear)
    & 0.171$\pm$0.060 & 21.51$\pm$3.41 & 0.709$\pm$0.099 & 0.205$\pm$0.059 & 0.337$\pm$0.083 & 28.75$\pm$8.38 \\
    IDW ($k{=}8$, $p{=}2$)
    & 0.180$\pm$0.057 & 20.94$\pm$2.92 & 0.728$\pm$0.089 & 0.252$\pm$0.076 & 0.397$\pm$0.099 & \textbf{27.11$\pm$8.06} \\

    \bottomrule
  \end{tabular}
\end{table}

\begin{figure}[t]
  \centering
  \includegraphics[width=\linewidth]{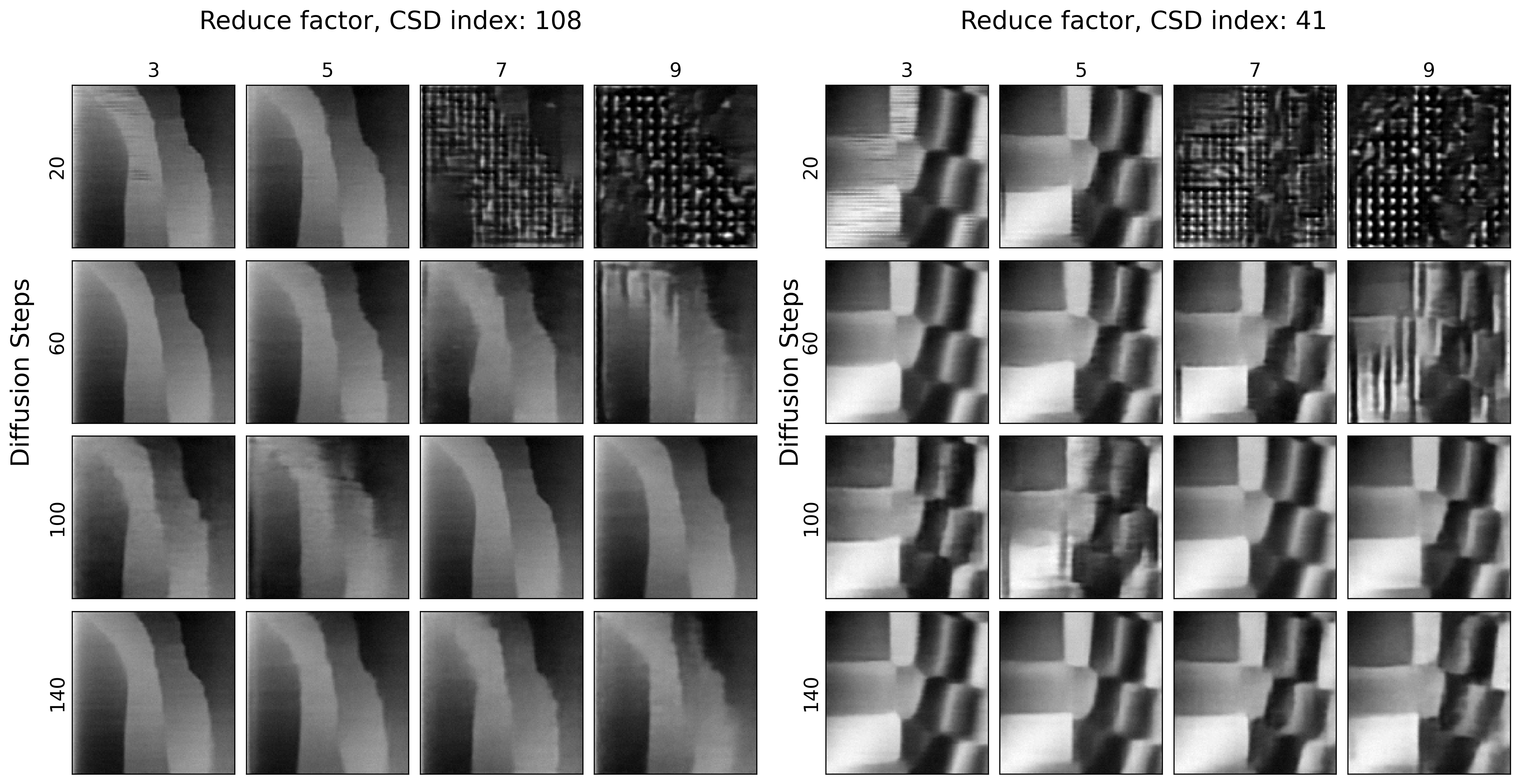}
  \caption{Reconstruction results for grid mask case, for reduce factor (3,5,7,9), using diffusion models with number of diffusion steps (20,60,100,140).}
  \label{diff_steps_reduced}
\end{figure}

\begin{figure}[t]
  \centering
  \includegraphics[width=\linewidth]{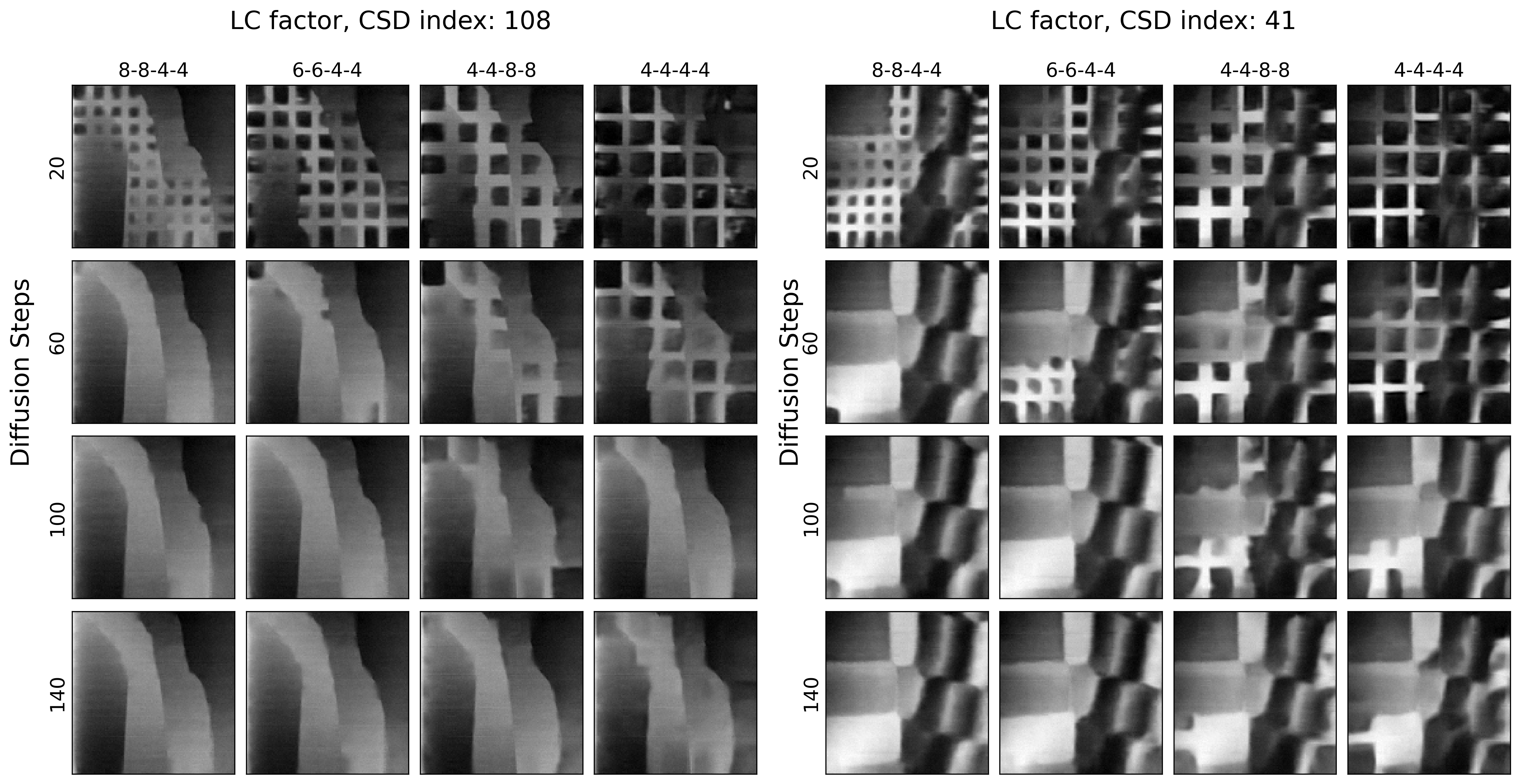}
  \caption{Reconstruction results for line-cut mask, for reduce factor (3,5,7,9), using diffusion models with number of diffusion steps (20,60,100,140).}
  \label{diff_steps_lc}
\end{figure}

\subsection{Time-Accuracy Trade-off}

For an experimental realization of the reconstruction pipeline proposed in this work, it is essential to minimize both the physical measurement time, and computational inference time. Figures \ref{diff_steps_reduced} and \ref{diff_steps_lc} show this relationship for the grid mask and the line-cut case, respectively, across different number of diffusion steps (20, 60, 100, 140) for the different sparsity level. This analysis is crucial because more diffusion steps generally improve quality but require longer inference time, creating a fundamental speed-accuracy trade-off.

For both masks, it is visible that to achieve high reconstruction quality for sparse measurements a lot of diffusion steps are needed. For grid masks with low reduce factor, the model succeeds in reconstructing the image even with as little as 20 diffusion steps, and for higher sparsity, 60 or 100 steps seem sufficient, which indicates an accuracy saturation with increasing inference time. For the line-cut case, the trend is similar but the improvement from 20 to 60 steps is even more accentuated. For the extreme sparsity case of LC factor (4,4,4,4), even using 140 diffusion steps seems to not yield optimal results, with masks artifacts still visible. This reinforces that line-cut reconstruction is a more difficult task, and requires a longer diffusion process to successfully bridge large unmeasured regions.

\begin{figure}[h]
  \centering
  \includegraphics[width=0.9\linewidth]{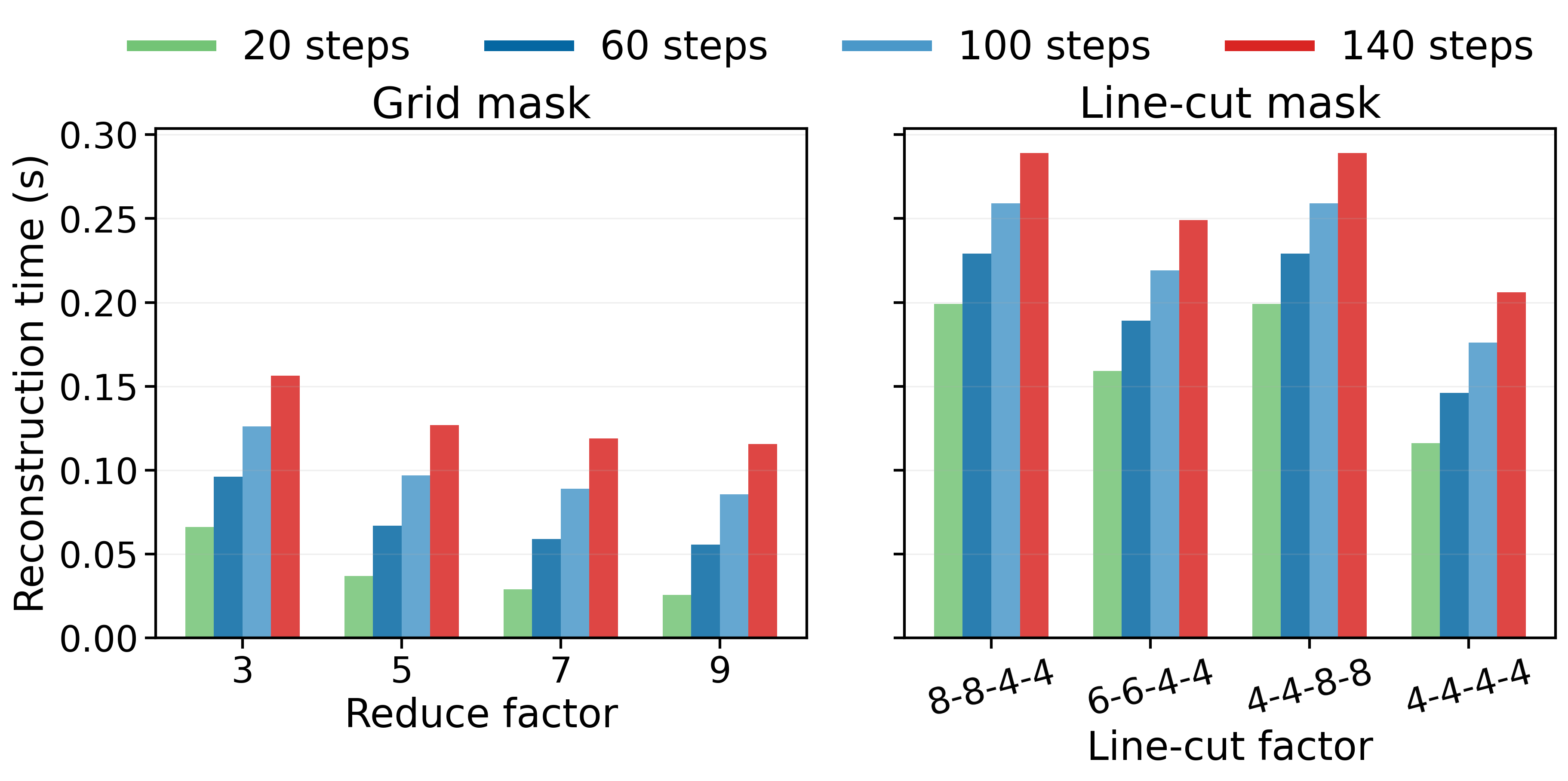}
  \caption{Estimated reconstruction time, adding measurement time with diffusion inference time, for reduce factor (3,5,7,9) in the case of grid mask (left panel), and (8-8-4-4, 6-6-4-4, 4-4-8-8, 4-4-4-4) in the case of line-cut mask (right panel).}
  \label{time}
\end{figure}

\begin{figure}[h]
  \centering
  \includegraphics[width=0.9\linewidth]{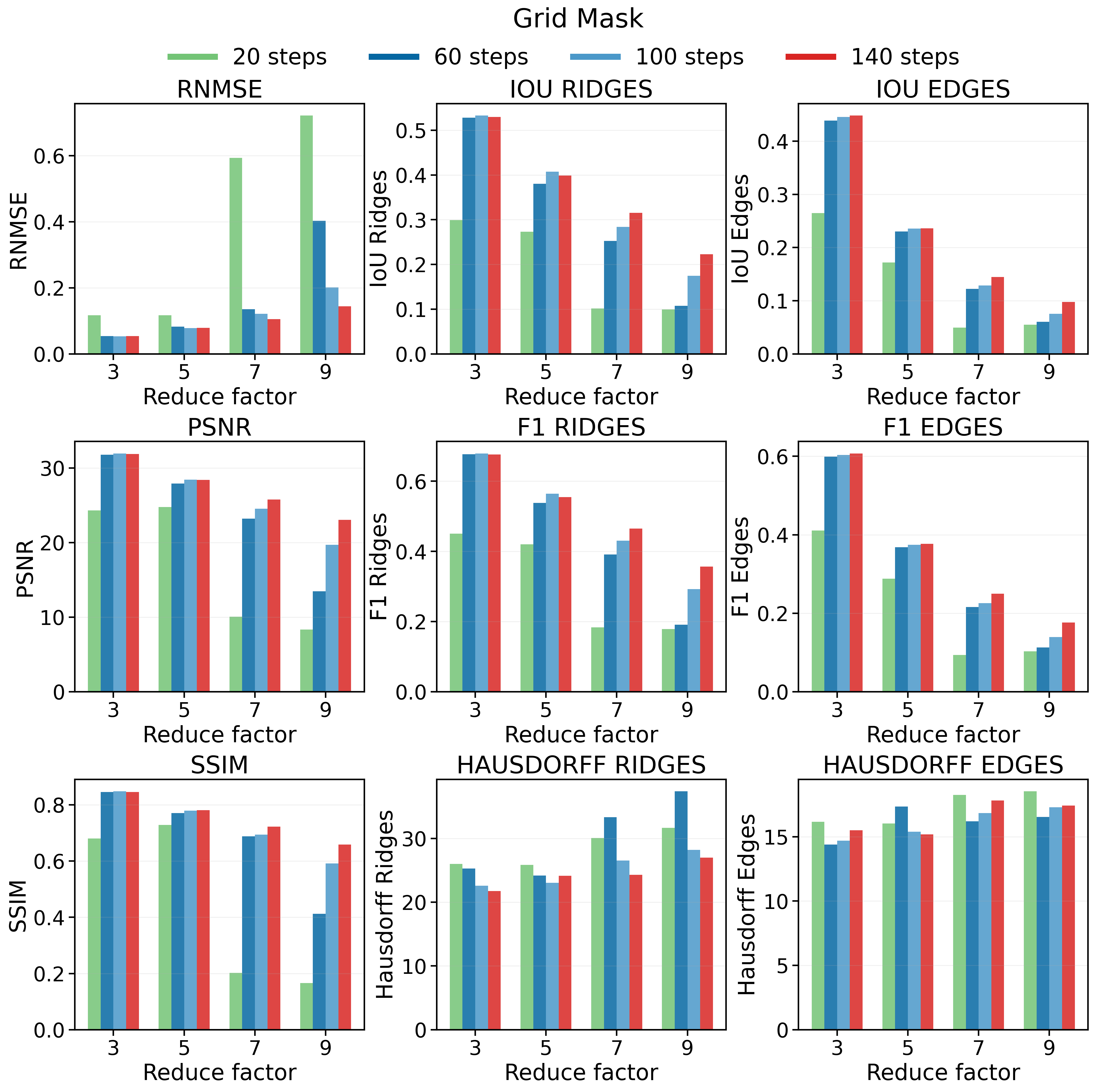}
  \caption{Reconstruction metrics for the grid mask, using reduce factor (3,5,7,9). Different number of diffusion steps are shown in different colors: green, dark blue, light blue, and red, for 20, 60, 100, and 140 diffusion steps, respectively.}
  \label{metrics_diffsteps_reduce}
\end{figure}

\begin{figure}[h]
  \centering
  \includegraphics[width=0.9\linewidth]{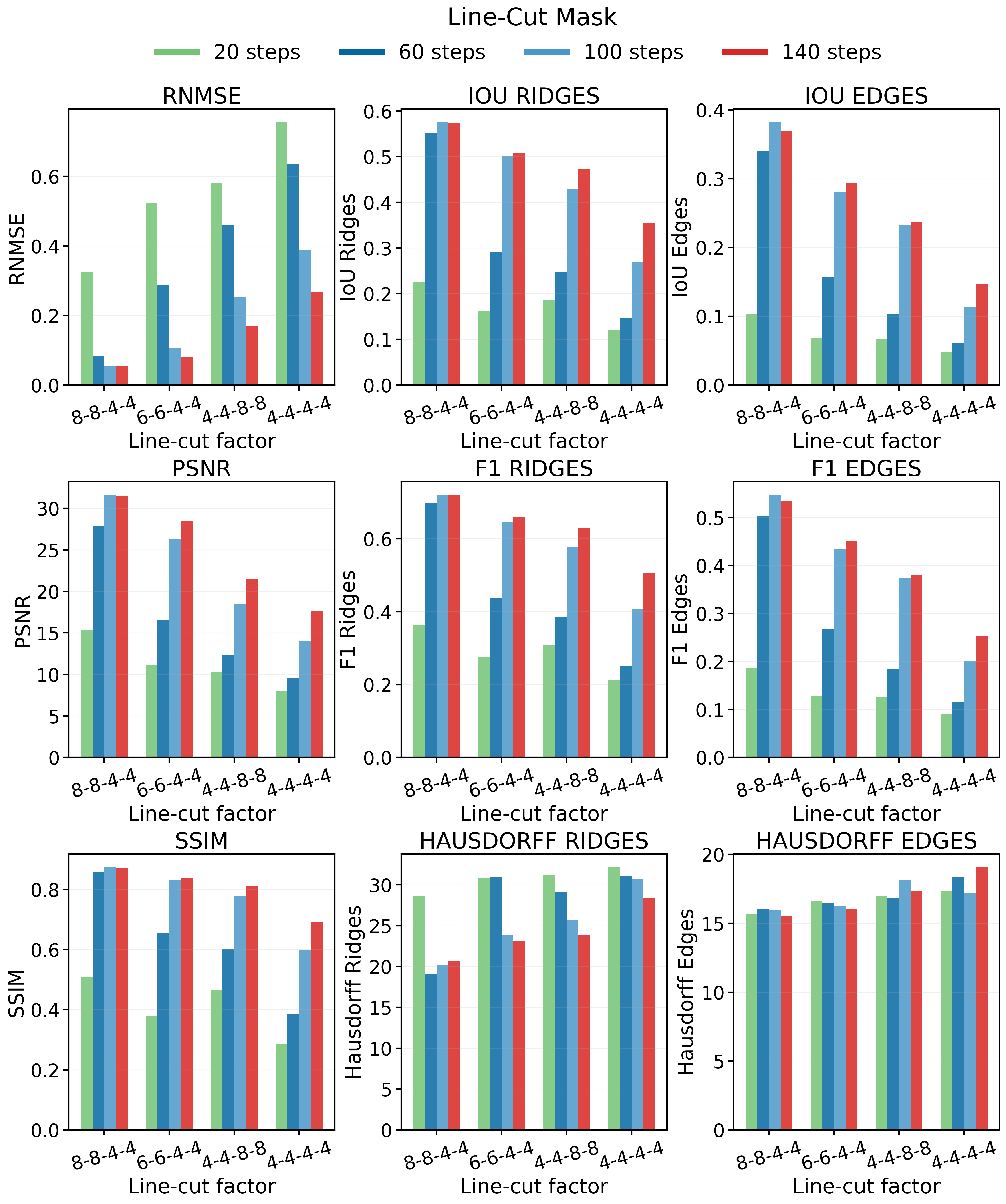}
  \caption{Reconstruction metrics for the line-cut mask, using line-cut factor (8-8-4-4, 6-6-4-4, 4-4-8-8, 4-4-4-4). Different number of diffusion steps are shown in different colors: green, dark blue, light blue, and red, for 20, 60, 100, and 140 diffusion steps, respectively.}
  \label{metrics_diffsteps_lc}
\end{figure}

Figures \ref{time}, \ref{metrics_diffsteps_reduce}, and \ref{metrics_diffsteps_lc} synthesize the relationship between mask sparsity, diffusion steps, and reconstruction fidelity, providing a complete map of the speed-accuracy trade-off. This analysis serves as practical guidance for experimentalists to select an ideal measurement protocol based on specific hardware constraints and desired throughput.

In Figure \ref{time}, we present the total \textit{time-to-reconstruct} a CSD for both grid (left) and line-cut (right) strategies across various sparsity levels. While the diffusion model's inference time remains approximately constant for a fixed number of steps, the total experimental overhead is dominated by the physical measurement phase. To provide a realistic benchmark, we calculate the time required for radio-frequency (RF) spectroscopy with a 25~$\mu$s integration time per pixel, representative of the high-speed acquisition methods currently utilized in the field. 

The total time required for acquisition and reconstruction is modeled as:
\begin{equation}
    T_{total} = (n_p \times t_p) + t_d,
    \label{time_eq}
\end{equation}
where $n_p$ is the number of measured pixels (determined by the mask configuration), $t_p$ is the integration time per pixel, and $t_d$ is the diffusion inference time on a dedicated GPU (e.g., approximately 0.02 s for 20 diffusion inference steps on a NVIDIA H-100). While this estimate represents an idealized lower bound by omitting system latencies such as data transfer between the experimental controller and the GPU, it remains a critical metric for comparing the efficiency of different masking protocols.

The relative speedup offered by our generative approach becomes even more pronounced when considering slower acquisition techniques, such as traditional DC measurements. Similarly, the measurement overhead per point is significantly higher when each pixel must be extracted based on repeated single-shot measurements instead of a direct charge measurement. This is the case in recent experiments where CSDs are reconstructed without direct access by charge sensors, and instead spin physics and spin shuttling are used to probe charge boundaries. In such regimes, the ability to reconstruct a complete diagram from a sparse scan (e.g., 4\% or 11\% measurement density) provides an order-of-magnitude reduction in device characterization time, directly addressing a primary bottleneck in quantum dot scaling and automated tuning.

From Figure \ref{time}, grid mask reconstruction achieves considerably shorter total CSD reconstruction times compared to line-cut masks, due to the combination of high sparsity and reduced measurement overhead, while inference time remains the same across mask types. While for line-cut, there is a significant reduction for sparser masks, such as LC factor (4,4,4,4), the grid mask case saturates very early, because for such low measurement points, the diffusion inference time starts to heavily dominate in Equation \ref{time_eq}.

Figures \ref{metrics_diffsteps_reduce} and \ref{metrics_diffsteps_lc} show reconstruction metrics for the grid mask and the line-cut mask, respectively, across reduce factors and diffusion steps, highlighting the accuracy saturation with increased inference time.

In the line-cut mask strategy, the metrics results for different sparsity levels and diffusion steps, shown in Figure \ref{metrics_diffsteps_lc} exhibit a similar trend to what was exposed in the previous section: reconstruction mainly works for diffusion steps $\geq$ 60, and low sparsity cases, like LC factors (8,8,4,4) and (6,6,4,4). 

For the grid mask case, metrics in Figure \ref{metrics_diffsteps_reduce} reveal that for reduce factors 3 and 5 accuracy saturates at around 60 diffusion steps, with further increasing being negligible, while considerably increasing inference time (Figure~\ref{time}). For sparser strategies (reduce factors 7 and 9) even models with 140 diffusion steps show significantly lower accuracy, across all metrics, when compared to reduce factors 3 and 5.

\section{Discussion}

\subsection{Why Diffusion Succeeds}

The success of our diffusion model compared to classical interpolation stems from its ability to learn and leverage global structural patterns from training data. Unlike interpolation methods that assume only local smoothness, our model has been exposed to approximately 9,000 diverse CSDs during training and has implicitly learned key patterns: charge transition lines are approximately linear, noise has characteristic patterns, and charge occupation regions are relatively uniform. This learned domain knowledge provides a powerful prior that enables plausible extrapolation far beyond what smoothness assumptions allow.

This is particularly evident in the line-cut results, where interpolation fails catastrophically despite measuring 23-44\% of pixels. The failure occurs because smoothness-based methods cannot bridge large contiguous gaps regardless of total measurement count, they require nearby measurements to interpolate between. In contrast, the diffusion model's learned structural prior enables it to generate plausible completions in these large unmeasured regions based on patterns it has seen during training. The key insight is that measurement spatial distribution is as crucial as measurement density: grid masks at 4\% density outperform line-cut masks at 44\% density because the former provides distributed conditioning throughout the image.

\subsection{Experimental Implications}

Our results demonstrate that diffusion-based reconstruction can provide meaningful acceleration of CSD acquisition. From the time-accuracy trade-off comparison, we highlight grid masking with reduce factor = 5, and 60 diffusion steps, as the best experimental strategy that balances speedup (5$\times$ compared to full measurement) with transition-line-reconstruction accuracy. 

Our training data spans diverse devices and noise levels. The consistent performance across test examples suggests the model has learned robust structural features rather than memorizing specific device configurations. However, we acknowledge important limitations: if a CSD exhibits radically different structure, the model may struggle. Deploying this approach in practice would require initial validation on new devices before relying on reconstructions for critical tuning decisions.

\section{Conclusion}

We have presented a conditional diffusion model for accelerating charge stability diagram (CSD) acquisition in quantum dots, through sparse measurement reconstruction. Our compact architecture, with approximately 160,000 parameters in a multi-scale U-Net, learns complex structural patterns from a dataset of 9,000 CSDs. The model successfully reconstructs transition lines from as little as 4\% of measured data.

The main insight enabling this performance is that diffusion models can learn and leverage global domain-specific structure, going beyond what is possible with classical interpolation, that rely on local smoothness. This advantage is particularly evident in line-cut masks, where interpolation methods fail to produce accurate reconstruction even with 23-44\% of measured context. This shows that the spatial distribution of measurements is as important as measurement density.

Promising directions for future work include online learning during experiments, exploration of alternative architecture (VAEs, GANs, CNNs), uncertainty estimation methods for targeted additional measurements, and integration with automated tuning algorithms. Beyond ML-oriented extensions, a promising direction for experimentalists is to use the diffusion model’s learned global structure to inform adaptive measurement strategies, identifying voltage regions that are most informative to sample next. This approach could reduce acquisition time while preserving accurate transition-line reconstruction, directly benefiting practical tuning workflows.

\section*{Acknowledgments}
We acknowledge useful discussions with Sam Katiraee-Far and Menno Veldhorst.

\section*{Funding}
This publication is part of the `Quantum Inspire -- the Dutch Quantum Computer in the Cloud' project (with project number NWA.1292.19.194) of the NWA research program `Research on Routes by Consortia (ORC)', which is funded by the Netherlands Organization for Scientific Research (NWO). This publication is also part of the project Optimal Digital-Analog Quantum Circuits with file number NGF.1582.22.026 of the research programme NGF-Quantum Delta NL 2022 which is (partly) financed by the Dutch Research Council (NWO) and the Dutch National Growth Fund initiative Quantum Delta NL. This project was supported by the Kavli Foundation. This work was also supported by the Netherlands Organization for Scientific Research (NWO/OCW), as part of Quantum Limits (project number SUMMIT.1.1016). R.K. and V.H. acknowledge support through the Horizon Europe Framework Programme's Integrated Germanium Quantum Technology (IGNITE) project under Grant Agreement No. 101069515, from the U.S. Army Research Office (ARO) under Award No. W911NF24-2-0043. R.K. acknowledges the KIND synergy program from the Kavli Institute of Nanoscience Delft.

\section*{Author contributions}
V.H., J.R., R.K., and T.S. designed the methodology, developed the code, performed the data analysis, and wrote the manuscript. B.U., A.C., and L.M.K.V. contributed to the conceptual development of the work through scientific discussions and provided critical feedback that improved the manuscript. E.G. conceived and supervised the project and contributed to writing and revising the manuscript.

\section*{Data availability statement}
The code used to produce the results and figures in this paper is publicly available on GitLab at \url{https://gitlab.com/QMAI/papers/diffusioncsds}. The dataset supporting this work is available on Zenodo at \url{https://doi.org/10.5281/zenodo.19252638}.

\bibliography{apssamp}

\end{document}